\documentclass{article}
\usepackage{arxiv}

\usepackage[utf8]{inputenc} 
\usepackage[T1]{fontenc}    
\usepackage{hyperref}       
\usepackage{url}            
\usepackage{booktabs}       
\usepackage{amsfonts}       
\usepackage{nicefrac}       
\usepackage{microtype}      
\usepackage{lipsum}
\usepackage{cite}
\usepackage{amsmath,amssymb,amsfonts}
\usepackage{algorithmic}
\usepackage{graphicx}
\usepackage{textcomp}
\usepackage{todonotes}
\usepackage{xcolor}
\usepackage{graphicx}
\graphicspath{ {./images/} }

\title{Surveying Off-Board and Extra-Vehicular Monitoring and Progress Towards Pervasive Diagnostics}

\author{
 Joshua E. Siegel \\
  Department of Computer Science and Engineering\\
  Michigan State University\\
  East Lansing, MI 48824 \\
  \texttt{jsiegel@msu.edu} \\
   \And
 Umberto Coda \\
 Department of Mechanical and Aerospace Engineering\\
 Politecnico di Torino\\
 Turin, Italy\\
 \texttt{umberto.coda@studenti.polito.it}
  \\
}

\begin{document}
\maketitle

\begin{abstract}
We survey the state-of-the-art in offboard diagnostics for vehicles, their occupants, and environments, with particular focus on vibroacoustic approaches. We identify promising application areas including data-driven management for shared mobility and automated fleets, usage-based insurance, and vehicle, occupant, and environmental state and condition monitoring. We close by exploring the particular application of vibroacoustic monitoring to vehicle diagnostics and prognostics and propose the introduction of automated vehicle- and context-specific model selection as a means of improving algorithm performance, e.g. to enable smartphone-resident diagnostics. Towards this vision, four strong-performing, interdependent classifiers are presented as a proof-of-concept for identifying vehicle configuration from acoustic signatures.  The described approach may serve as the first step in developing ``universal diagnostics,'' with applicability extending beyond the automotive domain.
\end{abstract}

\section{Motivating the Need for Vehicle Diagnostics}
The automotive world is changing, and there is increasing concern about vehicles' environmental impact, particularly those with internal combustion engines. One contributor to lifetime efficiency is vehicular diagnostics, as these systems may precisely report faults early, helping motivate owners and operators to seek out preventative or restorative maintenance. 

At the same time, mobility culture is evolving, transitioning from individual vehicle ownership towards mobility-as-a-service. Vehicle sales are slowing despite continued high mobility demand: the average vehicle age and lifetime miles travelled are increasing, particularly in developing countries~\cite{acea,bts}, and shared mobility services, car rentals, and ``\textit{robotaxis}'' are emerging. Increased utilization and novel use cases require enhanced fleet data generation and management capabilities. Automotive diagnostics, i.e. the inference of a vehicle's condition based on observed \textit{symptoms} indicating a technical state~\cite{cempel1988vibroacoustical}, are critical for effective fleet management. 

This survey considers the need for and approaches to vehicle diagnostics, first introducing technologies, capabilities, and shortcomings in traditional on-board approaches, and then considering the capabilities and advantages of newly-feasible off-board diagnostic systems. Specifically, we survey the state-of-the-art in offboard diagnostics for vehicles, their occupants, and environments, with particular focus on vibroacoustic approaches. We identify promising application areas including data-driven management for shared mobility and automated fleets, usage-based insurance, and vehicle, occupant, and environmental state and condition monitoring, and close by exploring the application of extravehicular diagnostics and prognostics and introduce the concept of automated vehicle- and context-specific diagnostic model selection as a means of improving algorithm performance in diverse and potentially-unknown fleets. Such a system could be used, for example, to enable smartphone-resident diagnostics, or in-garage, non-contact diagnostic tools. The described approach may serve as the first step in developing ``universal diagnostics'', with applicability extending pervasive sensing beyond the automotive domain and into factories, utilities, homes, healthcare, and beyond.

\section{Diagnostic Approaches}
Automotive diagnostics traditionally draw upon in-situ sensors and computation to support ``On Board Diagnostics,'' making use of data generated within a vehicle to diagnose the vehicle itself. Increasingly, extra-vehicular sensors - added on for diagnostic purposes, or present to enable other applications - may be used. In this section, we introduce and contrast the two approaches. 

\subsection{On-Board Diagnostics (OBD)}
On-Board Diagnostic (OBD-II) systems present on vehicles sold since 1996~\cite{8058008} are an automated control system utilizing distributed sensing across a vehicle's embedded systems as a technical solution for measuring vehicle operational parameters and detecting, reporting, and responding to faults.

Sensors may capture signals (e.g. vibration, or noise) and algorithms extract and process features, typically comparing these ``signatures'' against a library of previously-labeled reference values indicating operating state and/or failure mode~\cite{delvecchio2018vibro}. If a ``rule'' is triggered, an indicator is set to notify the user of the fault, and additional software routines may run to minimize the impact of the fault until the repair can be completed (e.g. by changing fuel tables). OBD data have also been used to enable indirect diagnostics, for example using the measured rate of change of coolant temperature to infer oil viscosity and therefore remaining useful life through constitutive relationships and fundamental process physics~\cite{6985355}. 

Certain OBD parameters are required to be reported by law in certain geographies. In some instances, there may be accuracy requirements. In others, parameters may not be reported or may be reported inaccurately. As a result, on-board diagnostics may not be accurate or effective.~\cite{defectsnotrecognizedbyOBD}

\subsection{Non-OBD Systems}
OBD is effective at detecting many fault classes, particularly those related to emissions~\cite{defectsnotrecognizedbyOBD}. However, some failure modalities may not be detected by OBD, or may be detected with slow response time or poor classification accuracy, because: 
\begin{itemize}
\item Incentive misalignment discourages the use of high-quality (costly) sensors, leading manufacturers to source the lowest-cost sensor capable of meeting legislative standards. Relying upon the data generated by these sensors leads to ``GIGO'' (Garbage In, Garbage Out)~\cite{siegel_ms_thesis_maybe}
\item Diagnostics may be tailored to under-report non-critical failures to improve customer satisfaction, brand perception, and reliability metrics relative to what might be experienced with an ``overly sensitive'' implementation
\item OBD systems are single-purpose, meaning they correctly identify the symptoms of the faults for which they were designed, but small performance perturbations may not be detected. For example, a system designed to enhance emissions may monitor engine exhaust gas composition continuously, but will not indicate wear or component failures leading to increased emissions until a legal threshold requiring occupant notification is surpassed~\cite{defectsnotrecognizedbyOBD}.
\end{itemize}

OBD's deficiencies are amplified by an ever-aging vehicle fleet~\cite{acea,bts}, though older cars stand to gain the most from the incremental reliability, performance and efficiency improvements enabled by adaptive and increasingly sensitive diagnostics. While newer vehicles may have the ability to update diagnostic capabilities remotely via over-the-air-updates~\cite{siegel_ota_patent}, older vehicles may lack connectivity or the computational resources necessary to implement these advanced algorithms. And while some diagnostic solutions may make use of manufacturer-proprietary data unavailable to OBD, particularly in newer and highly-sensored vehicles, this is not universally true. Further, the sensor payload in the incumbent vehicle fleet is immutable, with no data sources added post-production - that is, the sensors installed at time of sale are the sensors available at any point in the vehicle's life, and they are unlikely to get better with age. Therefore, the vehicles most in need of enhanced and robust diagnostics are the least-likely to support them. 

For these reasons, there is a need for updatable off-board diagnostics capable of sensitive measurement, upgradability, and enhanced prognostic (failure predictive) capabilities. A low-cost approach, even if imperfect, will enhance vehicle owners' and fleet managers' ability to detect, mitigate, and respond to faults, thereby improving fleet-wide safety, reliability, performance, and efficiency. 

\subsection{Diagnostic Opportunities in the Mobile Revolution}
As the need for enhanced fleet-wide utility grows, so to does the challenge of monitoring increasingly diverse vehicles and their associated, complex subsystems. The same enhancements driving the growth of in-vehicle sensing and connectivity have simultaneously empowered a parallel advance: namely, the growing capabilities of personal mobile devices. $70$\% of the world's population is now using smartphones~\cite{KANARACHOS2018867} possessing rich sensing, high-performance computation, and pervasive connectivity - capabilities enabling a diagnostic revolution. 

\subsubsection{Smartphone Sensing Capabilities}
While condition monitoring equipment has historically been cost-prohibitive, contemporary mobile devices include more sensors than ever, facilitating inexpensive and performant data capture with minimal complexity (see~\cite{kwapisz2011activity} for an example of pervasive mobile sensing as applied to human activity recognition). Initially, mobile device sensors served to support core device functions, with software libraries easing access to their data and widening their use cases into third-party applications~\cite{grossi2019sensor} and analytics. Today, mobile device sensor support continues to grow, and even older devices may add external sensors through serial, Bluetooth~\cite{GROSSI2019572} or Wi-Fi connectivity. Modern devices feature accelerometers, cameras, magnetometers, gyroscopes, GPS, proximity, light sensors and microphones that are accurate, precise, high-frequency, efficient and low-cost~\cite{KANARACHOS2018867,grossi2019sensor}. These capabilities have enabled the large-scale use of mobile systems as sensing devices in two-thirds of experimental research studies where such sensors are required~\cite{laport2019review}.  

The use of mobile devices as pervasive sensors has an added benefit of \textit{embodying intelligence}. That is, untrained users with access to the appropriate applications can make technically-sound judgments, identifying even those problems for which the device user has no prior knowledge of its existence -- and no awareness that the application is scanning for such faults~\cite{siegel_tire_cracks,siegel_afci}. This reduces the training burden for mechanics and fleet managers, and makes operating larger and more-diverse fleets feasible. 

By shifting intelligence from cost-, energy- and performance-constrained, in-vehicle hardware into third-party devices, the enabled algorithmic models may also be made more computationally-intensive, more easily updated, provided with access to higher-quality (and evolving) sensors and data, aggregable at a fleet level, and airgapped from critical vehicle hardware and software. 

This concept has been proven across domains. For example, introducing energy into a physical sample and studying the transient response across diverse sensors has been used to enable an individual to ``tap'' an object in order to determine its class~\cite{gong2019knocker}.

\subsubsection{Smartphone Computation and Connectivity Capabilities}
Pervasive connectivity enables diagnostics to utilize diverse data sources, and supports off-line processing and the creation of diagnostic algorithms capable of adapting over time. This is a result of having access to increased computational resources, enhanced storage capabilities, and richer fingerprint databases for classification and characterization. It also means that ``fault definitions'' may be updated at a remote endpoint, such that diagnostics may improve performance over time without requiring in-vehicle firmware upgrades (over-the-air or otherwise).

To this end, mobile phone computing power has recently increased, with the new mobile GPU Adreno 685~\cite{notebookcheck} reaching the computational power of Intel's 1998 ASCI Red supercomputer~\cite{businessinsider}. Networking capabilities have similarly grown, allowing for inexpensive global connectivity. 

While some vehicles offer connectivity~\cite{8058008} which may be used to support OBD's evolution, the use of a third-party devices has an additional benefit to manufacturers: with mobile devices, the users, not the manufacturer, pays for bandwidth and hardware capability upgrades over time. 

Mobile phones can augment or supplant the data generated by OBD, fusing in-vehicle sensing with smartphone capabilities to enable richer analytics. A framework for fusing multi-source information to return actionable information has been developed~\cite{wilfinger2013car}, and in another case, accelerometers have been used to improve OBD diagnostic accuracy and precision~\cite{makarova2018improvement}. They may even be used to enhance sensors' sampling rate to capture higher-frequency behavior reliably~\cite{yaqub2013smart}. 

Smartphones offer clear benefits over (or in conjunction with) on-board systems, particularly when constraints such as battery life, computation, and network limitations are thoughtfully addressed~\cite{KANARACHOS2018867}, and present a compelling enhancement over automotive diagnostics' ``business as usual'' by offering broader diagnostics with increased sensitivity, and the ability to improve over time - whether through model upgrades, or even federated learning approaches. Though individuals have long used their smartphones inside vehicles~\cite{mckinseymobility}, including plugged in and mounted, recent moves towards in-car wireless charging even more firmly establish mobile devices as incredibly powerful automotive sensing and compute devices with few constraints. 

\section{Off-board Smartphone-based Diagnostics}
Given the relatively poor performance of some on-board diagnostic systems and limited potential for future upgrades, there is an opportunity to use users' mobile devices as ``pervasive, offboard'' sensing tools capable of real-time and off-line vehicular diagnostics, prognostics, and analytics. The capabilities of such tools are growing and they may soon supplant on-board vehicle diagnostics entirely, moving diagnostics from low-cost OBD hardware, frozen at time of production, to performant, extensible, and easily-upgradable hardware and adaptive software algorithms capable of improving over time. The advantage of this approach goes beyond performance improvements to increase flexibility, enabling diagnostics that address any vehicle -- new or old, connected or isolated -- taking advantage of rich data collection, better-characterizable sensors, and scalable computing. 

Many effective ``pervasive'' sensing technologies revolve around the concept of remote sensing of sound and vibration utilizing onboard microphones and accelerometers, sensors core to mobile devices. This class of sensing is termed ``vibroacoustic sensing,'' as it captured vibration and acoustic emissions of an instrumented system. 

\subsection{Vibroacoustic Diagnostics}
Vibroacoustic diagnostic methods originate from specialists troubleshooting mechanisms based on sound and feel, dating back to well before the time of Steinmetz's famous (and perhaps apocryphal) chalk ``X''~\cite{steinmetz}. 

The vibroacoustic diagnostic method is non-intrusive, as sound can traverse mediums including air and ``open'' space and vibration can be conducted through surfaces without rigid mounting. It is therefore an attractive option for monitoring vehicle components~\cite{delvecchio2018vibro}. Experientially-trained mechanics may be highly-accurate using these methods, though there may be future specialist shortages~\cite{10.2307/26277743} leading to demand for automated diagnostics. 

There has been work to automate vibroacoustic diagnostics. Sound and vibration captured by microphones and accelerometers, for example, has been used as a surrogate for non-observable conditions including wear and performance level~\cite{cempel1988vibroacoustical}. Low-cost microphones have been used to identify pre-learned faults and differentiate normal from abnormal operation of mechanical equipment using acoustic features, providing a good degree of generalization~\cite{tossavainen2015sound}. Like sound (which itself is a vibration), vibration has been used as a surrogate for wear, with increasing intensity over time reasonably predicting time-to-failure~\cite{7438441}.  In fact, accelerometers have also been used to infer machinery performance using only vibration emissions as input~\cite{ginart2011smart}. Vibrational analysis may be coupled with OBD systems to improve diagnostic accuracy and precision,~\cite{makarova2018improvement}, or used in lieu of onboard measurements.

Vibroacoustics, counterintuitively, may be more precise than OBD because air-gaps provide a mechanism for isolating certain sounds and vibrations from sensors. While vibration may therefore be used to capture ``conductive'' time-series data, acoustic signals may be preferable in certain applications as the mode of transmission may serve to pre-condition input data and may transmit information related to multiple systems simultaneously~\cite{delvecchio2018vibro}. 

In some applications, mechanical vibration may be more informative than sound. An example is the classification of bearing operating states in an industrial environment using vibration signals along with rough sets theory for diagnostics, yielding high classification performance using analytical methods~\cite{NOWICKI1992141}.

Some diagnostic fingerprints are developed based on understanding of the underlying physical process, whereas others are latent patterns learned from experimental data collection~\cite{delvecchio2018vibro}. 

\subsubsection{Physics-based approach}
Real-world systems have inputs including energy, materials, control signals, and perturbation. It is possible to directly-measure inputs, outputs, and machine performance, but indirect measurement of residual processes (heat, noise, etc.) may be less-expensive and equally-useful diagnostically~\cite{cempel1988vibroacoustical}.

Vibration and sound are energy emissions stemming from mechanical interactions. Due to inherent imperfections, even rotating assemblies, such as gear meshes, may be modeled as a series of repeated impact events producing a characteristic noise or lateral motion~\cite{dkabrowski2016simultaneous}.

If one understands these processes, it becomes possible to model them and to engineer a series of features useful for system characterization. Modelling and processing techniques include \textit{frequency analysis}, \textit{cepstrum analysis}, \textit{filtering}, \textit{wavelet analysis}, among others. These generate features that are more-robust to small perturbations and therefore resistant to overfit when used in machine and deep learning algorithms. Other features describing waveforms may provide better discriminative properties. The features selected are informed by the engineer's knowledge of the physical process and what she or he believes likely to be informative in differentiating among particular states. Careful feature selection has the potential to improve diagnostic performance as well as reducing computation time, memory and storage requirements, and enhancing model generalizability. 

\subsubsection{Vibroacoustic Challenges}
Though VA is a compelling solution, it requires significant and diverse training data to achieve high performance and classification or gradation algorithms may be computationally-intensive and tailored to highly-specific systems. Accepting minimally-reduced performance to enhance algorithm generalizability and reduce computational performance, and/or shifting computation to scalable Cloud platforms, has the potential to make VA more powerful as a condition monitoring and preventative maintenance tool for vehicles and other systems. 

At the same time, smartphone processing power is increasing, and it may be possible to use a mobile device as a platform for real-time acoustic capture and processing, as demonstrated by Mielke~\cite{mielke2013smartphone}, and to do the same for vibration capture and analysis~\cite{7438441}.

Algorithms trained on few measurements may be inherently unstable, so multi-device crowdsourcing improves acoustic measurement classification confidence~\cite{vij2018smartphone}. Diverse, distributed devices lead to better training data and enhanced confidence in diagnostic results, though it is challenging to balance accuracy with system complexity~\cite{7328363} and to ensure samples represent usable input signals rather than background noise~\cite{delvecchio2018vibro}.

These challenges can be managed with careful implementation, helping pervasively-sensed VA attain strong performance when utilizing system-specific models for diagnostics and  proactive maintenance within automotive and other contexts. Example automotive applications include:
\begin{itemize}
\item Vehicle identification and component-level diagnostics (Section~\ref{vehicle})
\item Occupant and driver behavior monitoring and telemetry (Section ~\ref{occupant})
\item Environmental measurement and context identification (Section~\ref{environment})
\end{itemize}

In addition to these existent applications, we envision a possible means of improve VA performance through improved contextual-awareness, and describe this vision in (Section~\ref{vision}).

\subsection{Vehicle Condition}\label{vehicle}
Vehicles are increasingly complicated, though their mechanical embodiment typically comprises systems that translate and rotate, vibrating through use. There is a corpus of prior art focused on analysis of such systems. 

One example comes from Shen, et al., who developed an automated means of extracting robust features from rotating machinery, using an auto-encoder to find hidden and robust features indicative of operating condition and without prior knowledge or human intervention~\cite{SHEN2018170}. Mechanical systems wear down, leading to different operating states that a diagnostic tool must be able to detect in order to time preventative maintenance properly. To address this need, a "sound detective'' was created to classify the different operating states of various machines~\cite{derMauer2019}. 

Another approach to vibrational analysis utilizes constrained computation and embedded hardware. A Raspberry Pi was used to diagnose six common automotive faults using deep learning as a stable classification method (relative to decision trees), comparing four neural network architectures~\cite{lee2019comparative}.  It is unclear how these results generalize to other vehicle types and configurations, and whether they are less-sensitive to small data perturbations than other techniques. The use of a constrained system demonstrates the potential scalability of VA diagnostic approaches to mobile devices and those with similar capabilities. 

\subsubsection{Engine and Transmission}
Automotive engines, as with other reciprocating machinery, are difficult to diagnose because of the coupling among subsystems. Engines generate sound stemming from intake, exhaust, and fans, to combustion events, valve-train noise, piston slap, gear impacts, and fuel pumping. Each manifests uniquely and transmits across varied transmission-pathways, as examined in this comprehensive survey related to the use of vibroacoustic diagnostics for ICE's~\cite{delvecchio2018vibro}

For this reason, audio may be more suitable than vibration for identifying faults as the air-transmission path eliminates some system-coupling, making it easier to disaggregate signals~\cite{delvecchio2018vibro}.

It is difficult to select the appropriate degree of abstraction in generating reference features, and a highly-abstracted vibroacoustic emission model for diagnostics has been developed~\cite{komorska2013vibroacoustic}. In many studies, complete and accurate physical fault models are not available, so signal processing and machine learning techniques help improve classification performance. There are techniques for signal decomposition to better-highlight and associate features with significant engine events, and it is possible to guide classification tools through curated feature engineering including time-frequency analysis, or wavelet analysis~\cite{delvecchio2018vibro}.

Sensing engines can be done on resource-constrained devices and still enable continuous monitoring, with hardware-agnostic algorithm implementations~\cite{10.1145/1869983.1869992}. Another example used an Android mobile device to record vehicle audio, create frequency and spectral features, and detect engine faults by comparing recorded clips with reference audio files, where the authors could detect engine start, drive belt issues, and excess valve clearance~\cite{navearoysybingco}.

Engine misfiring is a typical within older vehicles due to component wear. Misfires have been detected in a contact-less acoustic method with $94$\% accuracy, relative to $82$\% accuracy attained from vibration signals. Without opening the hood and recording at the exhaust, the authors reached $85$\% classification accuracy from audio (which again outperformed vibration)~\cite{singh2019improved}. While some algorithms have been developed without physical process knowledge, others make use of system models to improve diagnostic performance. Use of aspects of the physical model help reduce algorithm complexity, requiring a feature engineering work before analysing the input data.

Siegel used feature extraction to reach $99$\% fault classification accuracy in another study of misfires, well exceeding the prior art. This work demonstrates that feature selection and reduction techniques based on Fisher and Relief Score are effective at improving both algorithm efficiency and accuracy, as well as the concept of ``Pareto Data'' -- data captured from low-quality sensors that have the potential to deliver high value when appropriately processed. In this case, data were collected from a comoddity smartphone microphone~\cite{siegel_misfire}. Similar acoustic data and engineered features have been successfully used to monitor the condition of engine air filters, helping to precisely time change events~\cite{siegel_air_filter} without the need for costly, high-fidelity, calibrated sensors.

Some feature engineering techniques, such as wavelet packet decomposition used in Siegel's misfire and air filter work, have found application in other engine diagnostic contexts such as identifying excessive engine valve clearance~\cite{figlus2014condition} and combustion events~\cite{10.1007/978-981-13-6577-5_13}. Other common faults relating to failed engine head gaskets, valve clearance issues, main gearbox, joints, faulty injectors and ignition components can also be detected thanks to vibrational analysis~\cite{komorska2013vibroacoustic}. Transmissions, too, may be monitored, and a damaged tooth in a gear can be diagnosed capturing sound and vibration at a distance~\cite{dkabrowski2016simultaneous}. Even high-speed rotating assemblies, such as turbochargers, can be monitored -- turbocharging is increasingly common to meet stringent economy and emissions standards, and engine compression surge has been identified and characterized by sound and vibration~\cite{marelli2019incipient}.

Non-automotive engines and fuel type can also be identified using VA approaches. Smartphone sensors were used to classify normal and atypical tappet adjustments of tractor engines with $98.3$\% accuracy~\cite{sasikumar2019non}, and fuel type can be determined based on vibrational mode -- with $95$\% accuracy~\cite{bkakowski2018vibroacoustic}.

Other studies have used physics to guide feature creation for indirect diagnostics, e.g. measuring one parameter to infer another. In~\cite{6985355} for example, the authors originally used engine temperature over time as a surrogate measure for oil viscosity and found promising results relating $\frac{dT}{dt}$ to viscosity. As it turns out, vibration may be used as further abstraction. By measuring engine vibration one may determine the engine speed (RPM) and it is possible to determine whether the car is in gear~\cite{wheel_imbalance} to identify when the car is at rest. As an extension of our previous work, we now note that using knowledge of the car's warm-up procedure (which typically involves a so-called ``fast idle'' until the engine warms up to temperature, to reduce emissions), is therefore possible to time how long it takes to go from fast idle (where the engine runs quickly to warm up and therefore reduce emissions) to slow idle and infer temperature from vibration, thereby creating a means of inferring oil viscosity from vibration alone and without the use of onboard temperature data. 

For the scope of minimize the knowledge gap between vehicle operators and expert mechanics, a mobile application called OtoMechanic has been designed. It uses sound to improve diagnostic precision relative to that of untrained users. Intelligence is embedded in a mobile application wherein a user uploads a recording of a car and answers related questions to produce a diagnostic result. The application works by reporting the label of the most-similar sample in a database as determined by a convolutional neural network (VGGish model). Peak diagnostic accuracy is $58.7$\% when identifying the correct class from twelve possibilities~\cite{morrison2019otomechanic}.

Algorithms have the most value when they are transferrable, as they can be trained on one system and applied to another with high performance. In one study, transferrability across similar engine geometries of different cars was considered in the context of detecting piston and cylinder wear, and measuring valve-train and roller bearing state~\cite{10.2307/26277743}.

Powertrain diagnostics are important, but it is equally important to instrument other vehicle subsystems. We look next to how offboard diagnostics have been applied to vehicle suspensions as a means of improving performance, safety, and comfort. 

\subsubsection{Wheel, Tire and Suspension}
As with powertrain diagnostics, suspensions may be monitored using vibroacoustic analysis, optical and other methods, or a combination of both.

In terms of VA, wireless microphones have been used to monitor wheel bearings and identify defects based on frequency-domain features~\cite{8430303}, and vibration analysis has been implemented o detect remaining useful life of mechanical components such as bearings~\cite{tang2019fault}. Similar data sources and algorithms have been exploited to identify the emergence of cracks in suspension beams~\cite{bialkowski2015early}.

Other VA approaches have been implemented, using accelerometers and GPS to measure tire pressure, tread depth, and wheel imbalance~\cite{siegel_tire_pressure,wheel_imbalance}, primarily using frequency-based features. Such solutions could be extended to instrumenting brakes, using frequency features and low-pass acceleration to measure specific pulsations occurring only under braking, or gyroscopes, to measure events taking place only when turning (or driving in a straight line). 

As noted earlier, researchers have demonstrated a means of diagnosing six vehicle component faults using vibration and Deep Learning Diagnostics algorithms running within constrained compute environments. Some of these diagnostics target wheels and suspensions, specifically at wheel imbalance, misalignment, brake judder, damping loss, wheel bearing failure, and constant-velocity joint failure. Each fault was selected as manifesting with characteristic vibrations and occurring at different frequencies. This research required vehicle to be driven at particular speeds in order to maximize signal. Accuracy varies, with a peak Matthew Correlation Coefficient of 0.994~\cite{8819436} - however, a small sample size and randomly-generated datasets with replacement may lead to overfit, artificially heightening the reported performance. 

Aside from accelerometers and GPS, other sensor measurands have been explored in the context of suspension diagnostics, with classification and gradation algorithms making use of sensors including mobile phone cameras. In one application, smartphone cameras were used to identify  tire degradation resulting from oxidation and cross-linking failures based on the appearance of characteristic patterns identifiable with a convolutional neural network~\cite{siegel_tire_cracks}. The novelty in this application was the concept of ``embedded intelligence,'' which took specialized knowledge (knowledge both that tires degrade over time, and the method through which degradation manifests and becomes visible) and built it into a tool deployable across hardware variants and requiring no training to operate effectively~\cite{siegel_tire_cracks}. The existence of the application itself made vehicle owners and operators aware of potential risks and fault modalities, and brought expert-level assessment to the hands of any user with a mobile device with a camera and internet connection. 

\subsubsection{Bodies / Noise, Vibration, and Harshness}
Recent studies have utilized MEMS accelerometers to investigate vehicle vibration indicative of vehicle body state and condition. Specifically, MEMS accelerometers allow the diagnosis of articulation events in articulated vehicles, e.g. buses. In one study, sensors were placed within the vehicle, with one located within each of the two vehicle segments in order to detect articulation events and monitor changes in bearing play resulting from wear and indicating a need for maintenance~\cite{szumilas2019mems}.

Vehicle occupants value fit and finish and a pleasant user experience while riding in a vehicle. To this end, there is an unmet need for realtime noise, vibration, and harshness (NVH) diagnostics. VA and other offboard techniques may find application in identifying and remediating the source of squeaks, rattles, and other in-cabin sounds in vehicles after delivery from the factory. 

\subsection{Vehicle Operating State}
Beyond monitoring vehicle condition and maintenance needs, offboard diagnostics have the potential to identify vehicle operating state in realtime, e.g. to identify whether a vehicle is moving or not, the position of the throttle, steering, or braking controls, or in which gear the selector is currently placed. To this end, mobile devices can be used to enable sensitive classification algorithms making use of accelerometers and cameras. 

At their simplest, mobile devices may be used to detect mode of transit~\cite{9347625}, such as whether someone is in a car and driving~\cite{servizi2019mining}. Some context-aware applications use sensor data to detect whether a vehicle is moving, and if so, to undertake appropriate actions and adaptations to enhance occupant safety, e.g. by disabling texting while in motion~\cite{aksamit2013adaptive}. The aforementioned study made use of accelerometers to supervise and eliminate false positive events from the training dataset, ultimately yielding a performance with $98$\% specificity and $97$\% sensitivity~\cite{aksamit2013adaptive}.

Others have used similar data to detect the operating state of a vehicle in order to identify lane changes or transit start- and end-points, using smartphones. The overall accuracy attained depends on the algorithm used and classification label, but ranges from $78.3$\% to $88.6$\% for one tree-bagging method~\cite{8924950}. 

Vehicle operating state is an ongoing area of research, with new developments exploring:
\begin{enumerate}
\item Accelerometer-based accident detection and response~\cite{7328363}, including one research project wherein smartphones were used to detect and respond to incidents taking place on all-terrain vehicles and capable of differentiating ``normal'' driving from simulated accidents with over $99$\% confidence~\cite{matuszczyk2016smartphone}. Some approaches use these data to automate rerouting~\cite{7328363}
\item Mobile phone cameras have been used to detect a vehicle's distance to leading traffic~\cite{10.1145/3287059}, providing realtime contextual information and situational awareness while affording older vehicles the benefits of modern (and typically expensive) advanced driver assistance systems
\item Using K-means clustering with acceleration data to identify driving modes, such as idling, acceleration, cruising, and turning as well as estimating fuel consumption~\cite{lehmann2017towards} (there are multiple methods for using mobile sensors as surrogate data to indirectly estimate fuel consumption)~\cite{KANARACHOS2019436}.
\end{enumerate}

This area of research is fast-evolving, particularly as context-sensitive applications gain prominence. Another fast-emerging application of pervasive sensing and offboard diagnostics is to occupant state and behavior monitoring. 

\subsection{Occupant Monitoring} \label{occupant}
Many automotive incidents resulting in injury or harm to property result from human activity. It is therefore essential to monitor not only the state and condition of a vehicle, but also to supervise the driver's state of health and attention in order to reduce unnecessary exposure to hazards and to promote safe and alert driving~\cite{aceable}.

Occupant monitor (including drivers and passengers) may be grouped broadly into three categories: 
\begin{enumerate}
\item Occupant \textit{State}, namely health and the capacity to pay attention to and engage with the act of driving
\item Occupant \textit{Behavior}, namely the manner of driving, including risks taken and other parameters informing telemetry, e.g. for informing actuarial models for insurers or for usage-based applications~\cite{siegel_ms_thesis_maybe}
\item Occupant \textit{Activities}, namely the actions taken by occupants within the vehicle (e.g. texting), with particular application to preventing or mitigating the effects of hazardous actions
\end{enumerate}

\subsubsection{Occupant State}
Vehicle occupant state may be monitored for a variety of reasons, e.g. related to drowsiness, drunkenness, or drugged behavior. Mobile phones have been used to detect and report drunk driving behavior, with accelerometers and orientation sensors informing driving style assessments indicative of drunkenness~\cite{7328363,5482295}, while other studies have shown the successful application of mobile device camera images to measuring occupant alertness. Drowsiness may also be monitored using smartphone data, helping to inform ADAS systems~\cite{khodairy2021driving}.

The main issue with occupant state is related to drunk driving state. With mobile phones placed in the vehicle there is the opportunity to detect that particular condition observing both the driving style~\cite{7328363} (using accelerometers and orientation sensors)~\cite{5482295} and the driver alertness monitoring the eye state with mobile device camera~\cite{8595428}. As with vehicle diagnostics, multiple sensor types may be used to monitor driver state~\cite{KANARACHOS2018867}. 

Counterintuitively, as highly automated driving grows in adoption, there will be growing demand for occupant metrics - at first, to ensure that drivers are ``safe to drive,'' and later, to make judgments as to how much to trust a driver's observations and control inputs relative to algorithms, e.g. to trust a lane keeping algorithm more than a drunk driver, but less than a sober driver. 

\subsubsection{Occupant Behaviors and Telemetry}
Smartphones have been widely deployed in order to develop telematics applications for vehicles and their occupants, using exterioceptive sensing to support ``off board supervision''~\cite{7891981,10.1145/2753497.2753535,ortiz2020vehicle}. These data have been used by insurance companies to monitor driver behaviors and to develop bespoke policies reflecting real-world use cases, risk profiles, and driver attitudes. 

Karanachos~\cite{KANARACHOS2018867} explores the performance of smartphone-derived data as it relates to algorithm performance, device capabilities, power consumption, positioning accuracy, and driver behavior, as applied to travel mode, time and routing, maneuvering, aggression, eco-friendliness, and reactiveness, all of which are critical to informing telemetry algorithms such as vehicle tracking or insurance. 

Pervasively-sensed data are used in three main insurance contexts, helping to:
\begin{enumerate}
\item Monitor a driver and/or vehicle's distance traveled, supporting usage-based insurance premiums~\cite{vavouranakis2017smartphone}. 
\item Supervise eco-driving~\cite{7328363}, using metrics such as vehicle use or driver behavior (including harshness of acceleration and cornering, with demonstrated performance achieving more than $70$\% accurate prediction~\cite{PAPADIMITRIOU201991}) to guide more-conservative behavior. Related to this, vehicle speed can be monitored with smartphone accelerometers alone, with an accuracy within $10$MPH of the ground truth~\cite{ustun2019speed}.
\item Observe driver strategy and maneuvering characteristics, to assess actuarial risk~\cite{7328363} and feed models with real-world data~\cite{vavouranakis2017smartphone} to inform premium pricing. This information may be used as input into learned statistical models representing drivers, vehicles, and mobile devices to detect risky driving maneuvers~\cite{7812763}. Notably, driving style and aggression level can be detected with inexpensive multi-purpose mobile phones~\cite{peng2018vehicle,johnson2011driving} and vehicles or drivers may be tracked to identify the  potential for high risk operation~\cite{guido2012estimation}, in cases with no additional sensors installed in the vehicle~\cite{6232298}. 
\end{enumerate}

Other behavior monitoring and telemetry use cases relate to safety, providing intelligent driver assistance by estimating road trajectory~\cite{peng2018vehicle}, using smartphones to measure turning or steering behavior (with $97.37$\% accuracy~\cite{8691707}), classifying road curvature and differentiating turn direction and type~\cite{8735446}, or offering even-finer measure of steering angle to detect careless driving or to enhance fine-grained lane control~\cite{chen2015invisible}. Some mobile phone data may identify driving events in order to inform path planning algorithms~\cite{alqudah2021machine}. In~\cite{8784284}, the authors were able to identify straight driving, stationary, turning, braking, and acceleration behaviors independently on the orientation of the device. These approaches may use several learning approaches, though many use end-to-end deep learning framework to extract features of driving behavior from smartphone sensor data. 

\subsubsection{Occupant Activity}
Human activity recognition has been widely studied outside vehicular contexts, and the performance of such studies suggest a likely transferrability to vehicular environments, with pervasive (ambient) or human monitoring gaining prominence. We consider in-vehicle and non-vehicular activity recognition in this survey, as the techniques demonstrated may inspire readers to reapply prior implementations or to adapt their methods to automotive contexts. 

In this study, we consider three categories of ``off-board'' sensing for human activity recognition. 
\begin{enumerate}
\item \textbf{In vehicle activity recognition:} Similarly to the use of pervasive sensing for drunk driver detection, mobile sensing has been applied to the recognition of non-driving behaviors within vehicles, for example distracted driving and texting-while-driving. Detecting texting-while-driving is based upon the observation of turning behavior, as measured by a single mobile device~\cite{8793032}. Mobile sensing solutions making use of optical sensors have also been demonstrated to detect driving context and identify potentially-dangerous states~\cite{10.1007/978-3-030-29513-4_11}. A survey of smartphone-based sensing in vehicles has been developed, describing activity recognition within vehicles including driver monitoring and the identification of potentially-hazardous situations~\cite{7328363}. 

\item{
\textbf{Workshop activity recognition:} Human-worn microphones and accelerometers have been used to monitor maintenance and assembly tasks within a workshop, reaching $84.4$\% accuracy for eight-state task classification with no false positives~\cite{10.1007/978-3-540-24646-6_2}. In another study, similar sensors were used to differentiate class categories included sawing, hammering, filing, drilling, grinding, sanding, opening a drawer, tightening a vice, and turning a screw driver using acceleration and audio data. For user-independent training, the study attained recall and precision of $66$\% and $63$\% respectively~\cite{1677514}. The methods demonstrated in identifying different work- and tool-use contexts may provide the basis for identify human engagement with various vehicle subcomponents, e.g. interaction with steering wheels, pedals, or buttons, helping create richer ``diagnostics'' for vehicle occupants and their use cases.
}
\item{
\textbf{General activity recognition:} Beyond identifying direct human-equipment interactions, mobile sensing has been applied to the creation of context-predictive and activity-aware systems~\cite{4487086}. Wearable sensors and mobile devices with similar capabilities have been used to detect user activities including eating, drinking, and speaking, with a four-state model attaining in-the-wild accuracy of $71.5$\%~\cite{10.1145/2370216.2370269}. In another study, user tasks were identified over a $10$-second window with $90$\% activity recognition rate~\cite{kwapisz2011activity}. In vehicles and mobile devices, computation is often constrained. Researchers have demonstrated activity classification using microphone, accelerometer, and pressure sensor from mobile devices in a low-resource framework. This algorithm was able to recognize 15-state human activity with 92.4\% performance in subject-independent online testing~\cite{doi:10.1155/2014/503291}.

Related to tailoring user experience, acoustic human activity recognition is an evolving field aimed at improving automotive Human Machine Interfaces (HMI) suitable across contexts. In one study, 22 activities were investigated and a classifier was developed reaching an $85$\% recognition rate~\cite{6343802}. Acoustic activity recognition may also be applied directly to general activity detection. 

In consumer electronics, activity or context recognition may be used to detect appliance use or to launch applications based on context, or used as sound labeling system thanks to ubiquitous microphones. Sound labeling and activity/context recognition helps augment classification approached by defining a context (environment) in order to limit the set of classes to be recognized before classifying an activity based on available mined datasets. In one sample application, $93.9$\% accuracy was reached on prerecorded clips with $89.6$\% performance for in-the-wild testing. The demonstrated system was able to attain similar-to-human levels of performance, when compared against human performance using crowd-sourcing service Amazon Mechanical Turk~\cite{laput2018ubicoustics} In~\cite{elizalde2018nels} human feedback is used to provide anchor training labels for ground truth, supporting continuous and adaptive learning of sounds. 
}
\end{enumerate}
Detecting activities within a vehicle - using acoustic sensing or other approaches - may help to tailor the vehicle user experience based on real-time use cases. Studying existing techniques for general activity recognition and applying this to an automotive context has the potential to improve the occupant experience as well as vehicle performance and reliability.


Of course, monitoring vehicles and their occupants alone does not yield a comprehensive picture of a vehicle's use case or context: the last remaining element to be monitored is the environment. 

\subsection{Environment}\label{environment}
Environment monitoring is a form of off-board diagnostic that may help to disaggregate ``external'' challenges from problems stemming from the vehicle or its use, e.g. in separating vibration stemming from cracks in the road from vibration caused by warped brake rotors. Environment monitoring is also a crucial step towards autonomous driving, helping algorithms understand their constraints and operate safely within design parameters.

Already, smartphones can be used as pervasive sensors capable of complementing contemporary ADAS implementations. In one study, vehicle parameters recorded from a mobile device accelerometer have been used to measure road anomalies and lane changes~\cite{gozick2012driver}. Vibroacoustic and other pervasively-sensed measurements have also been used for environment analysis. These may be used to calibrate ADAS systems by monitoring road condition, to classify lane markers or curves, to measure driver comfort levels, and as traffic-monitoring solutions. Some example pervasively-sensed environment monitoring approaches are described as follows: 

\begin{itemize}
\item Pavement \textbf{road quality} can be assessed by humans, though mobile-only solutions~\cite{10.1007/978-3-030-35543-2_20} may be lower-cost, faster, or offer broader coverage. Accelerometers may be used for detecting defects in the road such as potholes~\cite{7328363,sharma2019pothole,eriksson2008pothole,chugh2014road} or even road surface type (e.g. gravel detection, to adapt antilock braking sensitivity)~\cite{aleadelat2018estimation} or speed bump locations~\cite{kyriakou2021vehicles}. Road-surface materials and defects may also be detected from smartphone-captured images using learned texture-based descriptors~\cite{10.1007/978-3-319-91635-4_7}. It is also relevant to consider the weather when monitoring the road surface condition for safety, and microphone-based systems have demonstrated performance in detecting wet roadways~\cite{abdic2016detecting}. Captured at scale, smartphone data may be used to generate maps estimating road profiles, weather conditions, unevenness, and mapping condition more precisely and less expensively than traditional techniques~\cite{ZHAO201992,sedivy2019mechatronics}, with enhanced information perhaps improving safety~\cite{8166869}.  These data may be used to report road and traffic conditions to connected vehicles~\cite{siegel_sarma_patent_smart_message}.
\item \textbf{Curve} data and road classification may integrate with GPS data to increase the precision of navigation system. Mobile phone IMU's have been used to differentiate left from right and U-turns~\cite{8735446}, and it is reasonable to believe that combining camera images with IMU data (and LiDAR point clouds, if available), may help to generate higher-fidelity navigable maps for automated vehicles. 
\item The \textbf{comfort level} of bus passengers has been investigated with mobile phone sensors, attaining $90$\% classification accuracy for defined levels of occupant comfort~\cite{chinanalysis}.
\item Mobile sensing has been used to detect parking structure occupancy~\cite{cherian2019mobile}.
\item Acoustic analysis of \textbf{traffic} scenes with smartphone audio data has been used to classify the ``busyness'' of a street, with $100$\% efficacy for a two-state model and $77.6$\% accuracy for a three-state model. Such a solution may eliminate the need for dedicated infrastructure to monitor traffic, instead relying on user device measurements~\cite{vij2018smartphone}. In~\cite{7829341}, the authors implemented a 10-class model, classifying environments based on audio signatures indicating energy modulation patterns across time and frequency and attaining a mean accuracy of $79$\% after data augmentation. Audio may also be used to estimate vehicular speed changes~\cite{10.1007/978-3-319-61461-8_4}, and vibration may be, as well - using a convolutional neural network to estimate speed while eliminating the drift typically associated with double-integrating accelerometer data~\cite{9340328}
\item Offboard sensors lead many lives - as phones, game playing devices, and diagnostic tools - so it is important for devices to be able to identify their own \textbf{mobility use context}. One approach uses mobile device sensors and Hidden Markov Models to detect transit mode, choosing among bicycling, driving, walking, e-bikes, and taking the bus, attaining $93$\% accuracy~\cite{xiao2019detecting}, which may be used to create transit maps and/or to study individuals' behaviors~\cite{ahlstrom2019using}
\end{itemize} 

Though the described approaches relate primarily to cars, trucks, and busses, many solutions apply to other vehicles as well. Off-board diagnostics for additional vehicle classes are described below. 

\subsection{Non-automobile Vehicles}
Off-board and vibroacoustic diagnostics capabilities may be used for non-automotive, truck, or bus-type vehicles, including planes, trains, ships, and more: 
\begin{itemize}
    \item As with cars, \textbf{train} suspensions and bodies have been instrumented using vibroacoustic sensing. Train suspensions have been instrumented and monitored using vibrational analysis~\cite{chiou2019survey}. Brake surface condition has also been monitored with vibroacoustic diagnostics~\cite{sawczuk2016application}. Train bodies (NVH) have also been monitored, notably the doors on high-speed trains. Their condition may be inferred with the use of acoustic data~\cite{sun2019strategy}.
    \item \textbf{Aerial vehicle} propellers are subjected to high rotational speeds. If imbalanced or otherwise damaged, measurement of the resulting vibrations may lead to rapid fault detection and response~\cite{iannace2019fault}.
    \item In \textbf{maritime environments} vibroacoustic diagnostics has been implemented with the use of virtualized environments and virtual reality to allow remote human experts with access to spatial audio and body-worn transducers to diagnose failures remotely~\cite{barlow2019using}.
\end{itemize}

The applications for off-board, pervasive sensing and vibroacoustic diagnostics for system, environment, and context monitoring will continue to grow across vehicle classes. 

%



\section{A Vision for Pervasive Automotive Sensing}\label{vision}
There is an opportunity to continue the development of vibroacoustic diagnostics for automotive contexts, taking into consideration recent advances in pervasive sensing and emerging use cases for shared mobility and other high-touch, high-utilization use cases for automobiles. Such a solution may take advantage of mobile device sensors alone, or use such offboard sensors to compliment in-vehicle hardware. 

\subsection{Motivating The Need for Context-Specific Models}
Often, classification relies upon generalizable models to ensure the broadest applicability of an algorithm, perhaps at the expensive of performance. Occasionally, classifiers - such as activity recognition algorithms - may make use of ``personalized'' models. Personal Models are trained with a few minutes of individual (instance-specific) data, resulting in improved performance~\cite{weiss2012impact}. This approach may be extended from activity recognition to off-board vehicle diagnostics, with the creation of instance- or class-specific diagnostics algorithms. Selecting such algorithms will therefore first require the identification of the monitored instance or class, which is an ongoing research challenge. 

We propose the creation of a ``context-based model selection system,'' aimed at identifying the instrumented system precisely such that tailored models may be used for diagnostics and condition monitoring. 

Differentiating among vehicle makes, models, and use contexts will allow tailored classification algorithms to be used, with enhanced predictive accuracy, noise immunity, and other factors - thereby improving diagnostic accuracy and precision, and enabling the broader use of pervasive sensing solutions in lieu of dedicated onboard systems. 

There are grounds to believe that implementing such a system is feasible. Automotive enthusiasts can detect engine types and often specific vehicle makes and models from exhaust notes alone - and researchers have demonstrated success using computer algorithms to do the same, recording audio with digital voice recorders, extracting features, and testing different classifiers - finding that it is possible to use audio to differentiate vehicles~\cite{barai2014mechanical}. 

The more the application knows or infers about the instrumented system, the more accurate the diagnostic model implemented may become. 

\subsection{A Representative Implementation}
Though there are a multitude of ways in which to implement such a system, the authors have given consideration to several architectures and identified one promising path forward. The following subsections describe a representative implementation upon which a contextual identification system and model selection tool may be built in order to improve diagnostic accuracy and precision for vibroacoustic and other approaches.

The concept begins with the notion of Contextual Activation, i.e. the ability for a mobile device to launch a diagnostics application in background when needed, just as it might instead load a fitness app when detecting motion indicative of running.

With the application launched, sensor samples may be recorded, e.g. from the microphone and accelerometer. These data may then be used to identify the vehicle and engine category, perhaps classifying these based entirely on the noise produced, or in concept with additional data sources, such as a connected vehicle's Bluetooth address, its user/company's vehicle management database and so on.

Once the vehicle and variant is identified, this information may be used to identify operating mode, and from this, a ``personalized'' algorithm may be selected for diagnostic or other activities. 

In aggregate, the system might be imagined along the lines of a decision tree -- by selecting the appropriate leaf corresponding to the vehicle make, variant, and operating status, it becomes possible to select a similarly-specific prognostic or diagnostic algorithm tailored to the particular nuance of that system. Implemented carefully, the entire system may run seamlessly, such that the sample is captured, the context is identified, and the user is informed of issues worth her or his time, attention, and money to address. 

\subsubsection{Contextual Activation}
This seamlessness is key to the success of the proposed pervasive sensing concept -- to maximize the utility of a diagnostic application, it must require minimal user interaction. The use of contextual activation enables the application to operate data capture only when the mobile device is in or near a vehicle, and the vehicle is in the appropriate operating mode for the respective test (e.g. on, engine idling, in gear, or cruising at highway speeds on a straight road). This allows the software (built as a dedicated application inside the mobile device) to operate as a background task or to be launched automatically when the mobile device detects it is being used within an operating vehicle. 

Other potential implementations of this automatic, context-based software execution include automatic application launching when the phone is connected via Bluetooth to the car, or when a mapping or navigation application is opened. In this specific situation, the GPS and accelerometer may be utilized to understand the specific kind of road the vehicle is running on, as well as its speed, e.g. to disallow certain algorithms such as those used to detect wheel imbalance from running on cracked or gravel roads.

One possible embodiment of the system may comprise a ``context layer'' for generating characteristic features and/or uniquely-identifiable “fingerprints” for a particular system, which then passes system-level metadata (system type, other details, and confidence in each assessment), along with raw data and/or fingerprints to a classification and/or gradation system. This ``context layer'' may be used both in system training and testing, such that recorded samples may exist alongside related metadata and therefore allow for classification and gradation algorithms to improve over time, as increasing data volume generates richer training information even for hyper-specific and rare system configurations. 


The application may therefore capture raw signals and preprocess engineered features to be sent to a server (these fingerprints are space-efficient, easier to anonymize, more difficult to reverse, and repeatable), uploading these data at regular intervals. 

\subsubsection{Vehicle (and Instance) Identification}\label{vin}
The next step after identifying that the mobile device is in or near a vehicle will be vehicle identification, or identification of a grouping of similar vehicle variants. Depending on the system to be diagnosed, similarities may take place as a result of engine configuration, suspension geometry, and so on. 

A vehicle ``group'' may be identified by engine type - that is, configuration, displacement, and other geometric and design factors. For example, we may classify an engine to be gasoline powered, with an inline configuration, having 4 cylinders with $2.0$ liters of displacement, turbocharged aspiration, and manufactured by Ford.

If in our database we do not have any available diagnostic algorithm (e.g. a misfiring test~\cite{siegel_misfire}) for this engine type, we then look at increasingly less-specific parent class models, such as generic car-maker-independent gasoline I4 $2.0$ turbo engine. If this is also not available, we go higher- and higher-level until it is necessary to use the least-specific model, in this case, a model trained for all gasoline engines - at the cost of potentially-decreased model performance. Alternatively, we may consider to use a \textit{similar} engine, with slight difference in displacement or powered by LPG fuel. This process is shown in Figure~\ref{instance_prediction}.

\begin{figure*}[h!]
  \caption{A representative model selection process, indicating a means of identifying a vehicle variant and then selecting the most-specific diagnostic model available in order to improve predictive accuracy.}
  \includegraphics[width=\textwidth]{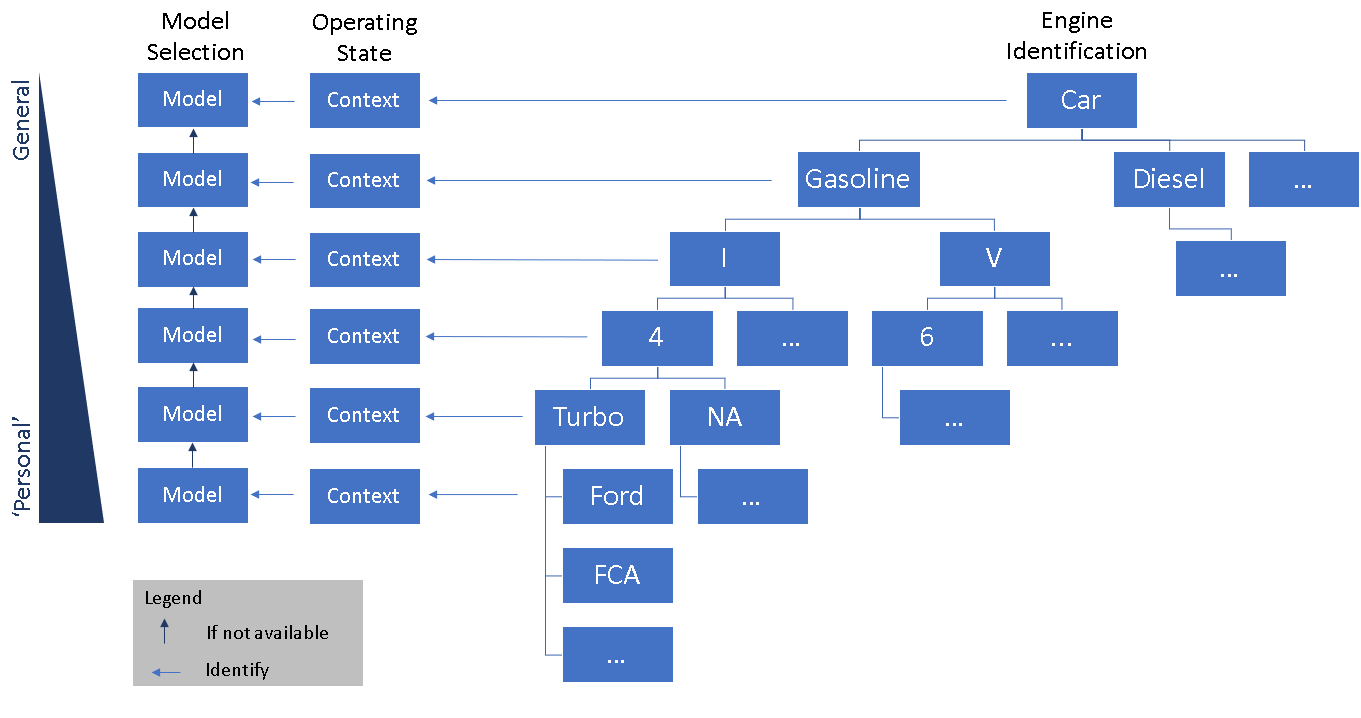}
\end{figure*}\label{instance_prediction} 

Extending this process, it may become possible to identify a particular vehicle instance, particularly based on features learned over time (e.g. indicating wear).

Other subsystems, such as bodies and suspensions, are harder to identify - but may still be feasible. For example, identifying operating context and road condition may be used to identify when a car hits a pothole, with the post-impact oscillations indicating the spring rate, mass, and damping characteristics indicative of a particular vehicle make or model. As with engines, subtleties may be used to identify vehicle instances, e.g. damping due to tire inflation. 

If the vehicle is known to the mobile device user and ``short list'' of vehicles frequented by the user, this portion of the classification may be replaced by ground-truth information, or selection may be made among a smaller/constrained subset of plausible options. Moreover, if we activate the application based on the Bluetooth connection indicating proximity to a particular vehicle, we may identify it with near-certainty. In order to reduce the degree of user interaction required, we may use this and other automation tools to identify vehicles and operating context in order to run engine and other diagnostics as a sort of background process. 

\subsubsection{Context Identification}
Once the vehicle is selected, its context must be identified. Context classification uses vibroacoustic cues (and vehicle data, if available) to identify the operating state of the engine, gearbox, and body. For example, is the engine on or off? If it is on, what is the engine RPM? Is the gearbox in park, neutral, or drive -- or if a manual transmission, in what gear is the transmission, and what is the clutch state?

Some algorithms will be able to operate with minimal information related to vehicle context (e.g. diagnosing poor suspension damping may require the vehicle simply to be moving as determined by GPS, whereas measuring tire pressure may require knowing the car is in gear~\cite{wheel_imbalance} and headed straight~\cite{siegel_tire_pressure} to minimize the impact of noise and other artifacts on classification performance. 

With context selection, we follow a similar process to that used for vehicle type and instance identification, selecting the model with metadata best reflecting the instrumented system to ensure the best fit and performance.

In an example implementation, we might create a decision tree to identify the current vehicle state - with consideration given to engine operating status, gear engagement, motion state, and other parameters - and rather than using this tree to select a model for diagnostics, we may prune this tree to suit a particular diagnostic application's needs (e.g. engine power might not matter for an interior NVH detection algorithm, or a tire pressure measurement algorithm may require the vehicle to be moving  to function~\cite{siegel_tire_pressure}. The pruned tree may then be used to select the ideal algorithm with the most-specific match between the training data and the current operating context.

With complicated vehicle operating contexts, and with systems measured under uncertainty, binary states may not be sufficient to describe the system status. For this reason, we instead propose the use of a three-state system comprising values of $-1$, $0$, and $1$.

If a context parameter is $1$, it is true or the condition is met. If it is $0$, it is false, or the condition is not met. If an identified context parameter is a negative value ($-1$) that means it is unnecessary for the diagnostic application, not available, uncertain, or not applicable (e.g. lateral acceleration is not applicable if a vehicle is stationary).

These negative values are removed from the input feature vector, and the corresponding element class is also removed from the reference database. In this way, a nearest neighbor matching algorithm will ignore uncertain or unnecessary data in considering the model to be used for diagnostics or prognostics. This matching algorithm needs a distance metric, which are algorithm-specific weighting coefficients used to define the importance of each context parameter (e.g. state of the engine may be more important than the amount of longitudinal acceleration when diagnosing motor mount condition, assuming both parameters are known).

A visual overview of the context identification and nearest-neighbor model selection process appears in Figure~\ref{context_identification}.

\begin{figure*}[h!]
  \caption{The model selection process relies on correct identification of both the vehicle variant and the context. Here, we see one proposed method for identifying the vehicle context and using those relevant features to select an appropriate "nearest neighbor" when identifying the optimal diagnostic or prognostic model to choose. Context parameters are identified through distinct, binary classifiers capable of reporting confidence metrics. The context vector comprises entries with three possible states (yes/no/uncertain or irrelevant), and those uncertain or irrelevant entries and their corresponding matches in the reference database are removed such that only confident, relevant parameters are used to select the nearest trained model.}
  \includegraphics[width=\textwidth]{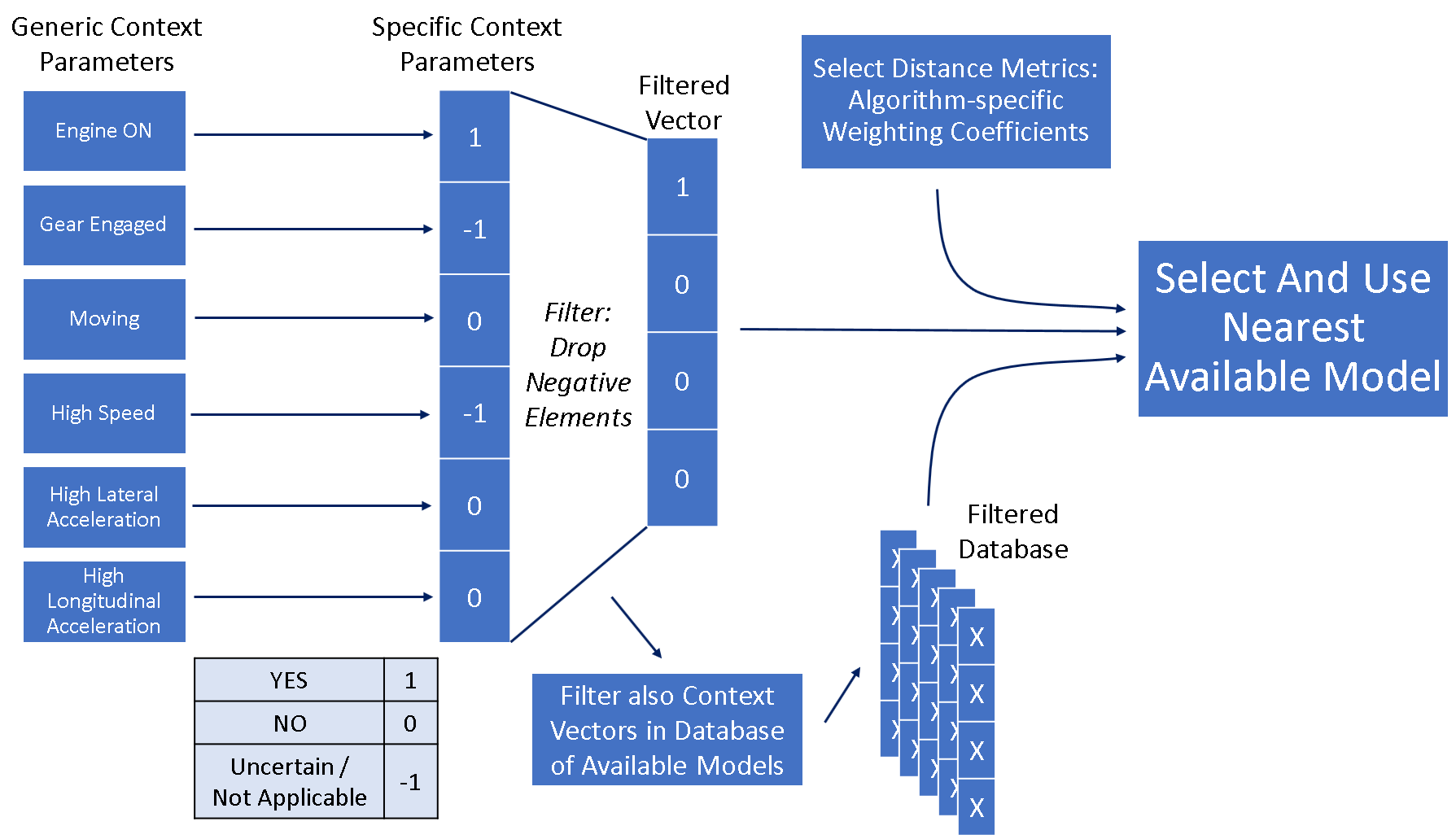}
\end{figure*}\label{context_identification}

Just as Bluetooth connectivity may be used to limit the plausible set of vehicle types, so too may data from sources such as on-board diagnostic systems be used to limit the set of feasible operating contexts, thereby removing uncertainty from the model selection process.

Combining vehicle identification with context classification, comprehensive vehicle ``metadata'' may be identified -- for example, ``light duty, 2.0 liter, turbocharged, Ford, Mustang, Joe's Mustang.'' With the fullest possible context identified, a list of feasible diagnostic algorithms may then be shortlisted.

\subsubsection{Diagnostics}
Certain diagnostics will be feasible for each set of vehicle classes and operating contexts. If a vehicle is moving, only algorithms working for moving vehicles will be available. If a vehicle is at idle, only algorithms operating at engine idle will be available. If a vehicle is on a gravel road, only algorithms suitable for rough terrain will be offered. 

When the mobile device identifies an appropriate context and short-lists feasible diagnostic algorithms, the most-specific diagnostic model of that type available with sufficient $n$ of training vehicles will be chosen and run on the raw data or engineered features provided by the mobile device (and vehicle sensors, if available). 

These algorithms will initially start out coarse - is the engine normal or abnormal? Are the brakes normal or abnormal?

Over time, as algorithms become more sensitive, and as training data are generated (with labeled or semi-supervised approaches), more classes may be added. The intent for this system is to transition from binary classification (good/bad), to gradation ($80$\% remaining life, $10$\% worn), to diagnostics so sensitive that they in fact are \textit{prognostics} -- that is, algorithms sensitive enough that faults may be detected and addressed proactively. 

The result will be improved efficiency, reliability, performance, and safety, and eased management of large-scale, high-utilization fleets, such as those that will be run by shared mobility services. The algorithms used may over time be adapted to minimize a cost function, e.g. balancing user experience with maintenance cost with the likelihood of having a car break down on the road. This will supplant data-blind proactive scheduled maintenance with data-driven insights sensitive to use environment, risk tolerance and mission-criticality.

\subsubsection{Preliminary Results}
First steps towards such a system are demonstrated in Coda's thesis (2020)~\cite{automobiles2020artificial}, that demonstrates the feasibility of developing classification systems to identify critical vehicle powertrain parameters useful for automated model selection through the creation of a flexible and user-friendly framework for testing varied featured generation and classification approaches. Feature extraction, machine learning, software framework, and results are presented for three different label categories: engine aspiration, fuel type, and cylinder count. These labels are predicted sequentially in order to exploit potential correlation, leading to a ROC-AUC higher than 93\% for the measured parameters in many cases. 

In this work, for which full details are provided in ~\cite{automobiles2020artificial}, samples of varied engines were recorded by the authors and their friends, from workshops, and from YouTube video clips of idling vehicles. Samples were captured variously from underhood, near a closed hood, and near the vehicle's exhaust. Data were manually labeled and in the case of uncertainty, labels were not assigned. Class balance was impacted by limited data availability, particularly reflecting a small number of ``Vee'' engines and low cylinder-count engines, though trends broadly reflected the imbalanced nature of real-world powertrain diversity. 

\begin{figure*}
	\centering
	\caption{The process by which captured engine audio is split into a set of informative features, as well as exploratory data analysis used to inform classifier design.}
	\includegraphics[width=0.8\textwidth]{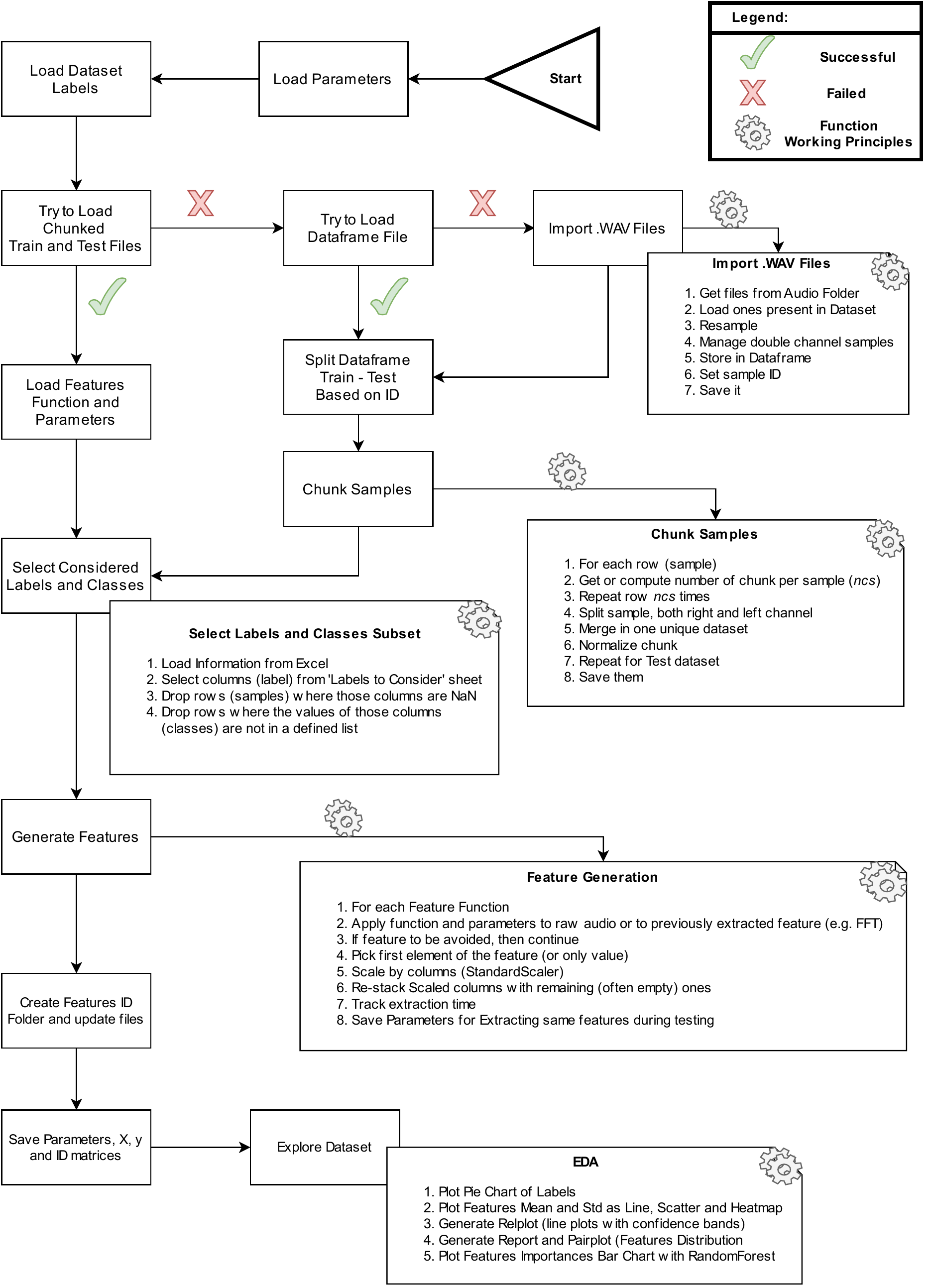}
	\label{CompleteToFeatures}
\end{figure*}

The work developed Python clip randomizer and feature extraction framework was developed to provide input into diverse classification models. Based on the success of Siegel's prior work~\cite{siegel_misfire,siegel_air_filter,wheel_imbalance}, a framework was created to support the generation of similar features including Fourier Coefficients, Mel-Frequency Cepstral Coefficients, and Discrete Wavelet Transform (DWT) features. These parameters captured critical waveform details that might be discernible to the human ear. In addition to these features, additional data such as skewness, kurtosis, power spectral density, and zero-crossing was also included to provide additional differentiating power. For each feature, the framework allowed for rapid configuration of feature parameters to aid in conducting a comprehensive grid search to find the globally-optimal model. Based on the results from exploratory data analysis, hypotheses were identified for testing within various classifiers. The data split and feature generation approach is shown in Figure~\ref{CompleteToFeatures}. 

\begin{figure*}
	\centering
	\caption{The process through which feature sets are loaded in order to test varied classification models.}
	\includegraphics[width=0.8\textwidth]{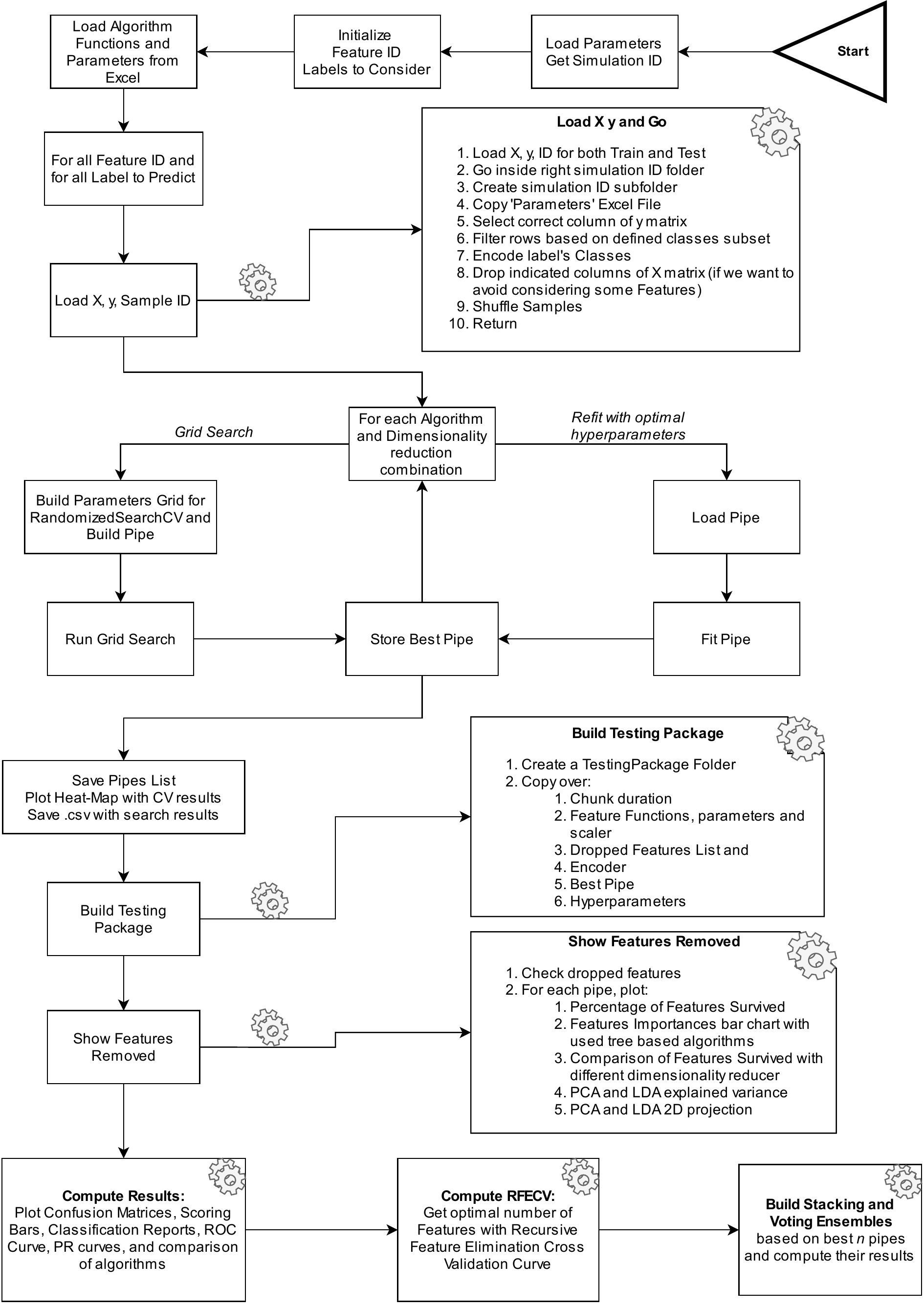}
	\label{overMainDiagram}
\end{figure*}

From the generated features, Coda implemented software to conduct a grid search over classifier models and hyperparameters, the flow of which is shown in Figure~\ref{overMainDiagram}.

In Coda's work, the first several context layers (Figure~\ref{instance_prediction}) are assumed - in this case, that the system is a light-duty vehicle, that and that it is idling. Aspiration type, fuel, and cylinder count are then classified as a means of working towards increasingly-specific diagnostic model selection. The ordering of this approach varies slightly to that proposed in Section~\ref{vin} and was determined based on apparent correlation among the sample dataset; the optimal ordering may be determined experimentally based on the available input data. 

From an exhaustive search described in the thesis, Coda found satisfactory performance for aspiration classification using an ExtraTrees classifier with Random Forest as a feature dimensionality reducer with the Receiver Operating Characteristic (ROC) Area Under the Curve (ROC-AUC)$=0.82$ and the Precision-Recall Area Under the Curve (PR-AUC)$=0.8$, with the confusion matrix in Table~\ref{turbo_table}. The classification result was dominated by Fast Fourier Transform (FFT) features, with some informative Mel-Frequency Cepstrum Coefficient (MFCC) features.

\begin{table}[]
\centering
\begin{tabular}{c|c|c|}
\cline{2-3}
                                                  & \textbf{Normally Aspirated} & \textbf{Turbocharged} \\ \hline
\multicolumn{1}{|c|}{\textbf{Normally Aspirated}} & 0.72                        & 0.09                  \\ \hline
\multicolumn{1}{|c|}{\textbf{Turbocharged}}       & 0.28                        & 0.91                  \\ \hline
\end{tabular}
\caption{This Confusion Matrix shows the results for Coda's aspiration-type classifier, which used an ExtraTrees classifier with Random Forest as a feature dimensionality reducer to classify based primarily upon FFT and MFCC features.}
\label{turbo_table}
\end{table}

Similarly, a fuel type classifier was developed using a grid search approach. In this case, aspiration status was used as an additional feature in determining fuel type. In this case, Gradient Boosting is the most-effective classifier, using FFT meta-statistics as input. In this case, Coda attained ROC-AUC ($0.99$) and PR-AUC ($0.994$), with the confusion matrix shown in Table~\ref{fuel_table}. Note that these results are for a single audio segment; if multiple segments are averaged and used to vote on the final classification; results improve further. 

\begin{table}[]
\centering
\begin{tabular}{c|c|c|}
\cline{2-3}
                                        & \textbf{Diesel} & \textbf{Gasoline} \\ \hline
\multicolumn{1}{|c|}{\textbf{Diesel}}   & 0.93            & 0.05              \\ \hline
\multicolumn{1}{|c|}{\textbf{Gasoline}} & 0.07            & 0.95              \\ \hline
\end{tabular}
\caption{This Confusion Matrix shows the results for Coda's fuel classifier, which used Gradient Boosting to classify primarily based on FFT meta-statistic features. }
\label{fuel_table}
\end{table}

Cylinder count was considered as the next and most-specific level of context for use in model selection within Coda's framework. This level of context classification entails multi-class labels, and suffers from class imbalance. While region-specific models may improve performance by excluding uncommon labels, broader models struggle to attain satisfactory performance. 

In the worst-case model, in which all available labels are represented, Coda found that using gradient boosting as a feature reducer and XGBoost as a classifier yielded the best performance, with a ROC-AUC=$0.93$ and PR-AUC=$0.856$. The confusion matrix appears in Table~\ref{cyl_table}.

\begin{table}[]
\centering
\begin{tabular}{c|c|c|c|c|}
\cline{2-5}
                                          & \textbf{3 Cylinder} & \textbf{4 Cylinder} & \textbf{6 Cylinder} & \textbf{8 Cylinder} \\ \hline
\multicolumn{1}{|c|}{\textbf{3 Cylinder}} & 0.5                 & 0.0072              & 0.24                & 0                   \\ \hline
\multicolumn{1}{|c|}{\textbf{4 Cylinder}} & 0                   & 0.82                & 0.5                 & 0.33                \\ \hline
\multicolumn{1}{|c|}{\textbf{6 Cylinder}} & 0.5                 & 0.072               & 0.12                & 0                   \\ \hline
\multicolumn{1}{|c|}{\textbf{8 Cylinder}} & 0                   & 0.099               & 0.14                & 0.67                \\ \hline
\end{tabular}
\caption{This Confusion Matrix shows the results for the worst-performing, broadest cylinder count classifier.}
\label{cyl_table}
\end{table}

Based on the strong per-parameter classification results for these three contextual classifiers, it is clear to see the feasibility in developing a suite of algorithms for determining a vehicle's use and operating context and how such data might be used as features in selecting an appropriate diagnostic model, whether for condition monitoring, or fault detection or analysis. With such a framework in place, it becomes feasible to select specific or generalized diagnostic algorithms based on the confidence of contextual classification and the availability of variously-tailored diagnostics. 

Though the authors are unable to share exact data due to commercial considerations, from the authors' own experience further developing the misfire work in \cite{siegel_misfire}, context-specific models tailored to a single engine variant demonstrate enhanced performance over a 15-vehicle trained generalized on the order of approximately 10\%, and better than a model trained on six vehicles with the similar engine configuration by approximately 5\%.

Worth noting is that the data used were from varied, uncalibrated devices. While some suggest that calibration may be necessary to attain quantitative, rather than qualitative, results~\cite{Manka2021}, it was not found to be necessary when using appropriately pre-processed data to differentiate among vehicle configurations.  

\subsection{Future Enhancements}
This solution paints a bold picture for the future related to transitioning on-board diagnostic systems into off-board, consumer-owned devices with the potential to upgrade both software and hardware over time. However, it is the first step in a more significant undertaking designed to revolutionize automotive diagnostics and maintenance, particularly for ride-share companies already reliant upon mobile applications for driver accessibility and vehicle tracking.

As part of a mobile application, customers may soon report data about vehicle health back to the fleet manager, and their phone will be used to collect data and to pay for the bandwidth of the logger. Beyond mobile devices, vibroacoustic diagnostics might soon be built into garage door openers, service stations, or parking lots. 

Future enhancements will utilize diagnostic results in conjunction with emerging technologies such as augmented and virtual reality and 3D printing to guide component inspection, maintenance, production, and replacement, with AR helping walk untrained users through component inspection, testing, and replacement, even guiding them through validating diagnoses. The same mobile devices used for diagnostics may then be used to access Augmented and Virtual Reality visualizations of components and their wear states or fault conditions~\cite{malaguti2014augmented}. Connected vehicle services~\cite{8058008} may be used to automate repair and maintenance scheduling, to minimize downtime for shared fleets. 

This concept is exciting in an automotive context, but the real value lay in applications farther afield. If context identification and model are proven out, the concept of pervasive sensing and vibroacoustic diagnostics may extend more broadly into ``universal diagnostics,'' wherein the same techniques discussed in this manuscript and particularly in the vision section may be used for ubiquitous sensing of other device and system types and classes. More than just cars, trucks, motorcycles, and bicycles may be monitored. Later, appliances including washing machines and microwaves. 

Already, some of this work has been done -- though without our proposed multi-step framework to pick hyper-specialized models based on device type and use context. Some of this work uses vibroacoustic analysis applied across the automotive life-cycle, from monitoring production equipment~\cite{8231733} to measuring process outputs to estimating (and automatically improving) the condition of automotive subsystems~\cite{cempel1988vibroacoustical}.

Other vibroacoustic signals have been used to diagnose faults in other electromechanical systems, such as power tools and coffee grinders~\cite{glowacz2019}. Similar techniques may help instrument people, diagnosing early-onset Parkinson's disease~\cite{ellis2015validated}. 

Combining pervasive sensing with enhanced diagnostics and embodied intelligence - that is, the ability for an application to bring expert knowledge to non-expert users - has the potential to change the world well beyond computer science, revolutionizing mechanical, chemical, and electric engineering, materials, science, and beyond. 

Our hope is that the survey provided and vision presented is a call to action for talented researchers to explore this green field, applying the technology of today to build - and maintain - the world of tomorrow more effectively. 

\subsection{Acknowledgements}
Umberto Coda's thesis work, which developed the presented proof-of-concept vehicle context identification system, was supported under a scholarship provided by Fiat Chrysler Automobiles (FCA) (now Stellantis). FCA was not involved with concept development, research methodology, or data analysis for this system. FCA was not involved with the survey coverage contained in this manuscript. 


\bibliographystyle{unsrt}  
\bibliography{diagnostics}
\end{document}